\documentclass[aps,prstab,reprint,twocolumn,groupedaddress,showpacs,amsmath,amssymb]{revtex4-1}
\usepackage{epsfig}
\usepackage{lipsum}
\usepackage{amsmath}
\usepackage{dcolumn}
\usepackage{bm}
\usepackage{esint}
\usepackage{xcolor}

\begin{document}


\title{Particle beam eigen-emittances, phase integral, vorticity, and rotations}


\author{L.~Groening and C.~Xiao}
\affiliation{GSI Helmholtzzentrum f\"ur Schwerionenforschung GmbH, Darmstadt D-64291, Germany}
\email[]{la.groening@gsi.de}
\author{M.~Chung}
\affiliation{Ulsan National Institute of Science and Technology, Ulsan 44919, Republic of Korea}
\email[]{mchung@unist.ac.kr}


\date{\today}

\begin{abstract}
Particle beam eigen-emittances comprise the lowest set of rms-emittances that can be imposed to a beam through symplectic optical elements. For cases of practical relevance this paper introduces an approximation providing a very simple and powerful relation between transverse eigen-emittance variation and the beam phase integral. This relation enormously facilitates modeling eigen-emittance tailoring scenarios. It reveals that difference of eigen-emittances is given by the beam phase integral or vorticity rather than by angular momentum. Within the approximation any beam is equivalent to two objects rotating at angular velocities~$\pm\omega$. A description through circular beam modes has been done already in [A.~Burov, S.~Nagaitsev, and Y.~Derbenev, Circular modes, beam adapters, and their applications in beam optics, Phys. Rev. E {\bf 66}, 016503 (2002)]. The new relation presented here is a complementary and vivid approach to provide a physical picture of the nature of eigen-emittances for cases of practical interest.
\end{abstract}


\maketitle


\section{Introduction}
The terms of eigen-emittances have been introduced by A.J.~Dragt in 1992~\cite{Dragt} as the projected rms-emittances a beam acquires after all correlations between the degrees of freedom (planes) have been removed. Accordingly, they form the set of lowest projected beam emittances which can be achieved by applying symplectic and linear beam line elements. Since this set may not fit requirements of certain beam applications, tailoring of eigen-emittances became a subject of extensive theoretical and experimental research. The first proposal of eigen-emittance modification was made in~\cite{Brinkmann_rep} followed by other fundamental investigations~\cite{burov_pre_2002,Kim,Bertrand,Emma_prstab2006,Groening_prstab2011,Chao_prstab2011,carlsten_prstab2011may,carlsten_prstab2011aug,duffy,Mccrady,Xiao_prstab2013,Xiao_nim,Groening_arxiv} and experimental applications in linear electron~\cite{Edwards,Brinkmann_prstab,Piot_prstab2006,Sun_prl2010} and ion accelerators~\cite{groening_prl,appel_nim_2017,chung_prl}. Measurements of eigen-emittances and beam coupling have been reported in~\cite{prat_prstab,oegren_prstab,xiao_nim_U,xiao_prab_ROSE,cathey_prl,xiao_nim_ROSE} for instance.

Albeit the underlying theory is well understood and experimental results match analytical calculations and layouts, the physical nature of eigen-emittances lacks a picture being more vivid compared to the common sense definition of "projected rms-emittances after removal of inter-plane correlations". For some special cases simple relations were derived as for beams with cylindrical symmetry dominated by their angular momentum, for which the difference of the two eigen-emittance is equal to the angular momentum and their mean is the transverse rms-emittance~\cite{Kim}.

This general lack of tangible comprehension is partially due to the fact that already the two transverse eigen-emittances are calculated in the four-dimensional (4d) phase space spanned by two space coordinates and two momentum coordinates, not to mention the three eigen-emittances of a beam including also longitudinal coordinates. This paper intends to shrink this lack of vivid understanding at least for the two transverse eigen-emittances.

In the next section basic terms are defined and the derivation of the extended Busch theorem is revised and adapted for further use. The third section derives the role of beam vorticity and the phase integral for transverse eigen-emittances. Section~\ref{sec_approx} introduces an approximation that is valid for cases of practical interest. It shows that in general it is vorticity being relevant for the difference of eigen-emittances rather than angular momentum. Afterwards several applications demonstrate the practical power of the beam phase integral on eigen-emittance calculation. The sixth section shows that eigen-emittances can be taken practically as equivalents of two areas rotating with same angular velocity but with opposite signs. The paper closes with a conclusion and an outlook.

\section{Basic terms and extended Busch theorem}
\label{s_ebt}
The two transverse eigen-emittances $\varepsilon_{1/2}$ originally introduced in~\cite{Dragt} are equal to the two projected transverse beam rms-emittances $\varepsilon_{x/y}$, if and only if there are no correlations between the two transverse planes. Eigen-emittances can be obtained by solving the complex equation
\begin{equation}
det(JC-i\varepsilon_{1/2}I)\,=\,0\,,
\end{equation}
where $I$ is the identity matrix,
\begin{equation}
\label{2nd_mom_matrix}
C=
\begin{bmatrix}
\langle x^2 \rangle &  \langle xx'\rangle &  \langle xy\rangle & \langle xy'\rangle \\
\langle xx'\rangle &  \langle x'^2\rangle & \langle yx'\rangle & \langle x'y'\rangle \\
\langle xy\rangle &  \langle yx'\rangle &  \langle y^2\rangle & \langle yy'\rangle \\
\langle xy'\rangle &  \langle x'y'\rangle & \langle yy'\rangle & \langle y'^2\rangle
\end{bmatrix}\,,
\end{equation}
and
\begin{equation}
\label{e_JMatrix}
J=
\begin{bmatrix}
0 &  1 &  0 & 0 \\
-1 &  0 &  0 & 0 \\
0 &  0 &  0 & 1 \\
0 &  0 & -1 & 0
\end{bmatrix}\,,
\end{equation}
where $(x,y)$ denote the particle position coordinates and $(x',y')$ their derivatives $(x',y')$ w.r.t.~the longitudinal direction $\vec{s}$. Second moments $\langle uv\rangle $ are defined through a normalized distribution function $f_b$ as
\begin{equation}
\langle uv\rangle\:=\,\int\int\int\int f_b(x,x',y,y')\cdot uv\cdot dx\,dx'\,dy\,dy'\,.
\end{equation}
The two transverse eigen-emittances can be calculated as~\cite{Xiao_prstab2013}
\begin{equation}
\label{eigen12}
\varepsilon_{1/2}=\frac{1}{2} \sqrt{-tr[(CJ)^2] \pm \sqrt{tr^2[(CJ)^2]-16\,det\,C) }}\,.
\end{equation}
Projected transverse beam rms-emittances are defined as~\cite{Floettmann_prstab}
\begin{flalign}
\label{emrms_x}
&\varepsilon_x^2\,=\, \langle x^2 \rangle\langle x'^2\rangle\,-\,\langle xx' \rangle ^2\,,\\
\label{emrms_y}
&\varepsilon_y^2\,=\, \langle y^2 \rangle\langle y'^2\rangle\,-\,\langle yy' \rangle ^2 \,.
\end{flalign}

\begin{figure}[hbt]
\centering
\includegraphics*[width=88mm,clip=]{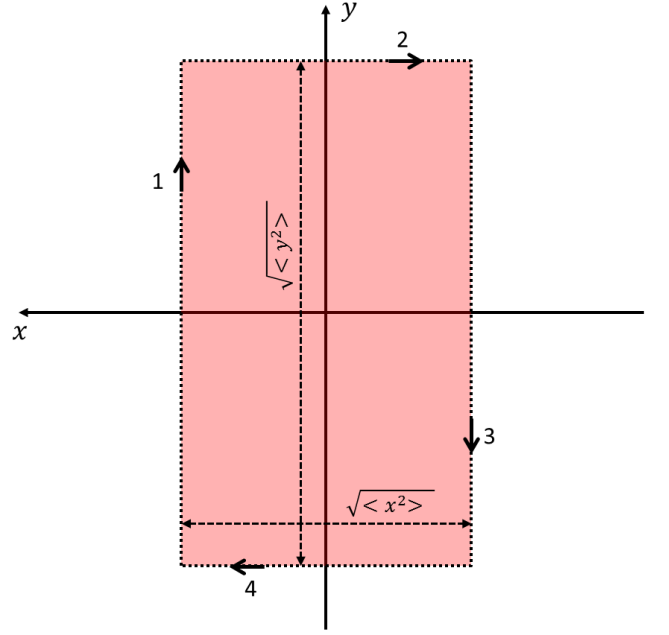}
\caption{Definition of the beam rms-area for the case of \mbox{$\langle xy\rangle =0$} and of the four path sections $P_i$ comprising the integral enclosing this area.}
\label{fig_vorticity}
\end{figure}

Busch's original theorem is from stating preservation of longitudinal single particle angular momentum in a magnetic field region with cylindrical symmetry using conjugated momentum~\cite{Busch}. The extended theorem is derived from re-stating preservation of many-particle beam eigen-emittances under symplectic transformations using conjugated momenta~\cite{groening_prab}. It reads
\begin{equation}
\begin{split}
\label{const}
& (\varepsilon_1-\varepsilon_2)^2\,+\,\left[\frac{AB_s}{(B\rho)}\right] ^2\,+\\
& \,2\frac{B_s}{(B\rho)}\left[\langle y^2\rangle\langle xy'\rangle - \langle x^2\rangle\langle yx'\rangle + \langle xy\rangle (\langle xx'\rangle - \langle yy'\rangle)\right]\,\\
& =\, const\,,
\end{split}
\end{equation}
where
\begin{equation}
\label{e_Arms}
A\,:=\,\sqrt{\langle x^2\rangle\langle y^2\rangle -\langle xy\rangle ^2}
\end{equation}
is the rms-area of the beam (Fig.~\ref{fig_vorticity}), $B_s$ is the longitudinal magnetic field along the beam axis, and $(B\rho)$ is the beam magnetic rigidity.
The first term of Eq.~(\ref{const}) is the squared difference of eigen-emittances and the second term is the square of the magnetic flux through the beam rms-area~$\vec{A}$.
Within the subsequent section the meaning of the essential part of the last term
\begin{equation}
\label{rms_Vorticity}
\mathcal{W}A\,:=\,\langle y^2\rangle\langle xy'\rangle - \langle x^2\rangle\langle yx'\rangle + \langle xy\rangle (\langle xx'\rangle - \langle yy'\rangle)
\end{equation}
is re-derived by adapting the method reported in \cite{groening_prab} to further use.

\section{Beam vorticity and phase integral}
\label{s_beam_vort}
This section starts with proving the relation
\begin{equation}
\label{e_Wa}
\mathcal{W}\,=\,\int\limits_A \,\left[\vec{\nabla}\times\vec{\bar{r'}}(x,y,s)\right]\cdot d\vec{A}\,,
\end{equation}
where $\vec{\bar{r'}}(x,y,s) :=[\bar{x'}(x,y,s),\bar{y'}(x,y,s),1]$. This relation states that the third term of Eq.~(\ref{const}) is mainly the mean rms-vorticity $(\vec{\nabla}\times\vec{\bar{r'}})$ integrated over the beam rms-area.

\begin{figure}[hbt]
\centering
\includegraphics*[width=88mm,clip=]{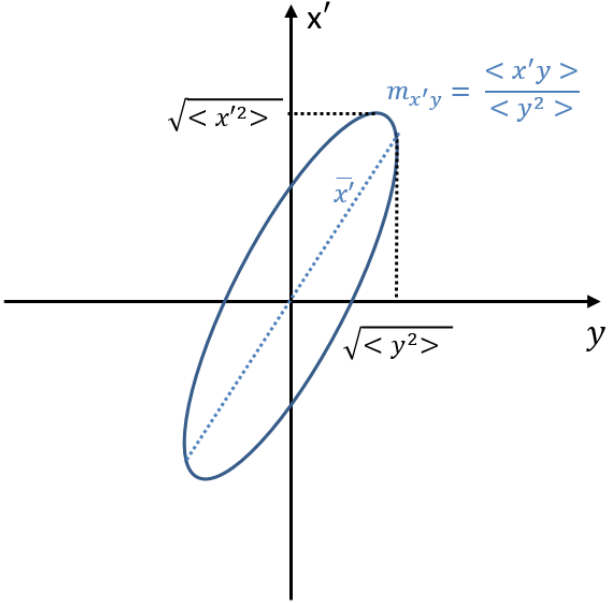}
\caption{Projection of the four-dimensional rms-ellipsoid onto the $(y,x')$~plane and the constant slope $\partial \bar{x'}/\partial y$.}
\label{xsy_plane}
\end{figure}

By construction, $\mathcal{W}$ from Eq.~(\ref{e_Wa}) is invariant under rotation. Hence, Eq.~(\ref{e_Wa}) can be expanded for $\langle xy\rangle =0$ without loss of generality (by assuming that prior to determination of $\mathcal{W}$ the beam is rotated around the beam axis by an angle that puts $\langle xy\rangle$ to zero~\cite{rot_inv}). Figure~\ref{fig_vorticity} illustrates this beam rms-area and the phase integral enclosing it.
The transverse components of $\vec{\bar{r'}}$ are expressed through
\begin{align}
\label{mean_rs_a}
\bar{x'}(x,y)\,:=\,\frac{\langle x'x\rangle }{\langle x^2\rangle }x\,+\,\frac{\langle x'y\rangle }{\langle y^2\rangle }y,\\
\bar{y'}(x,y)\,:=\,\frac{\langle y'x\rangle }{\langle x^2\rangle }x\,+\,\frac{\langle y'y\rangle }{\langle y^2\rangle }y.
\label{mean_rs_b}
\end{align}
Figure~\ref{xsy_plane} illustrates as an example the constant slope $\partial \bar{x'}/\partial y$ of $\bar{x'}$ in the projection of the four-dimensional rms-ellipsoid onto the $(y,x')$~plane.
Accordingly, 
\begin{equation}
|\vec{\nabla}\times\vec{\bar{r'}}|\,=\,\frac{\langle y'x\rangle}{\langle x^2\rangle}-\frac{\langle yx'\rangle}{\langle y^2\rangle}
\end{equation}
not explicitly depending on~($x,y$) hence turning the integral into simple multiplication with $A$ (using $\langle xy\rangle$=0) and to
\begin{equation}
\begin{split}
A\mathcal{W}\,&=\,A^2|\vec{\nabla}\times\vec{\bar{r'}}| \\
&=\,\langle x^2\rangle\langle y^2\rangle\,|\vec{\nabla}\times\vec{\bar{r'}}|\,=\,\langle y^2\rangle\langle y'x\rangle - \langle x^2\rangle\langle yx'\rangle \,,
\end{split}
\end{equation}
which had to be proven.
Using Stoke's theorem the quantity $\mathcal{W}$ is introduced as
\begin{equation}
	\mathcal{W}\,=\,\int\limits_A \left[\vec{\nabla}\times\vec{\bar{r'}}\right] \cdot d\vec{A}\,=\,\oint\limits_\mathcal{C}\vec{\bar{r'}}\cdot d\vec{C}
\end{equation}
being the beam phase integral. The r.h.s. of the equation refers to the integral enclosing the beam rms-area.

Equation~(\ref{const}) is accordingly rephrased as
\begin{equation}
\label{const_resorted}
(\Delta\varepsilon)^2\,+\,2\frac{AB_s}{(B\rho)}\mathcal{W}\,+\,\left[\frac{AB_s}{(B\rho)}\right]^2\,=\,const\,,
\end{equation}
with $\Delta\varepsilon:=(\varepsilon_1-\varepsilon_2)$, being the extended Busch theorem re-formulated using the beam phase integral. For constant beam rigidity this expression is equivalent to
\begin{equation}
	\begin{split}
		\label{d_sum_eq_0}
		&\delta((\Delta\varepsilon )^2)\,+\,2\frac{\mathcal{W}}{(B\rho)}(A\delta B_s + B_s\delta A)\\
		&+2\frac{AB_s}{(B\rho)}\delta W\,+\,2\frac{AB_s}{(B\rho)^2}(A\delta B_s + B_s\delta A)\,=\,0\,.
	\end{split}
\end{equation}

\section{Approximations in real applications}
\label{sec_approx}
The extended Busch theorem as stated above is exact. This section illustrates that the actual circumstances of experimental scenarios allow to derive powerful and simple relations from the theorem.

Experimental applications at beam lines with limited appertures keep as short as possible the sections along which eigen-emittances shall be modified. This is from the fact that these modifications require imposed correlations, which blow up projected emittances and beam sizes. Accordingly, the beam rms-area along such short sections can be approximated as constant~($\delta A$=0). Secondly, the assumption is made that for the above motivated practicality, behind such sections the beam is again fully uncorrelated, i.e., $\mathcal{W}$ vanishes according to Eq.~(\ref{rms_Vorticity}). Together with the above prerequisites the assumption is made that change of vorticity is strongly dominated and basically given by the change of the longitudinal magnetic field
\begin{equation}
\label{dv_eq_dW}
-\delta\mathcal{W}\,:=\,A\,\frac{\delta B_s}{(B\rho)}\,,
\end{equation}
being equivalent to the statement
\begin{equation}
\label{v_p_W_const}
\frac{AB_s}{(B\rho)}\,+\,\mathcal{W}\,:=\,const\,.
\end{equation}
Plugging this statement into above Eq.~(\ref{d_sum_eq_0}) delivers \mbox{$\delta ((\Delta\varepsilon )^2)=\delta (\mathcal{W}^2)$} and accordingly the simple and useful relation
\begin{equation}
\label{square_diff_const}
(\Delta\varepsilon )^2-\mathcal{W}^2\,=\,const\,.
\end{equation}

This relation states that variation of difference of eigen-emittances is given by variation of the beam phase integral being equal to the integrated beam vorticity. It is in line with the fact that short tilted quadrupoles and dipoles, which do not change the vorticity, do not change the eigen-emittances either. But solenoid fringe fields impose beam rotation, i.e., vorticity and do change the beam eigen-emittances.

Throughout this paper the validity of Eqs.~(\ref{v_p_W_const}) and (\ref{square_diff_const}) is assumed under the prerequisites stated above, i.e., constant beam rms-area and complete decoupling shortly after tailoring of eigen-emittances has been accomplished. The relevance of the latter assumption becomes clear from the formulation of $\mathcal{W}$ as phase integral: if there remains some coupling and the beam passes a long drift, $\mathcal{W}$ might change since $\vec{r'}$ remains constant along the drift but the rms-area~$A$ varies in shape and size. Eigen-emittances in turn do not change along any drift. Accordingly, the assumption of full decoupling after intended eigen-emittance modification is essential for the application of~Eqs.~(\ref{v_p_W_const}) and (\ref{square_diff_const}) after the decoupling. However, the intended tailoring process itself is decribed correctly by the relations.

In the following the contents of Eqs.~(\ref{v_p_W_const}) and (\ref{square_diff_const}) are benchmarked by applying them to several scenarios. They are proofed to full generality for cylindrical symmetric beams in solenoids dominated by their angular momentum.

\section{Applications}
\label{s_examples}
This section performs applications and hence testing of the relations stated above. First, the relations' properties w.r.t.~symplecticity are investigated followed by calculation of eigen-emittances of simple objects, the exact description of cylindrical symmetric beams, a simulation of a non-symmetric beam scenario, and successful modeling of experiments on eigen-emittance tailoring performed at FERMILAB and at GSI.

\subsection{Symplecticity}
The first check of Eq.~(\ref{square_diff_const}) is on verification whether symplectic transformations leave invariant the vorticity as they leave invariant eigen-emittances. Assuming a given transport matrix $M$ the absolute amount of vorticity will change through $M$ by
\begin{equation}
\label{D_vort}
\delta |\vec{\nabla}\times\vec{r'}|\,=\,\frac{d}{dx}\delta y'-\frac{d}{dy}\delta x'\,=\,m_{41}-m_{23}\,.
\end{equation}
The property $m_{41}-m_{23}$=0 is fulfilled by commonly used optical elements being linear and symplectic in the full 4d~transverse phase space, as drifts, dipoles, quadrupoles, solenoids, rf-gaps, and even by (non-symplectic) homogeneous central field regions inside of a solenoid~\cite{safe_ass}, as they feature~$m_{41}$=$m_{23}$=0.

Instead, fringe fields of solenoids are not symplectic, i.e., $m_{41}$=-$m_{23}\neq$0. They trigger additional beam vorticity of
\begin{equation}
\label{eq_rot_dB}
\delta |\vec{\nabla}\times\vec{r'}|\,=\,m_{41}-m_{23}\,=\,-\frac{\delta B_s}{(B\rho)}\,,
\end{equation}
where $\delta B_s$ is the change of field strength along the fringe region.
Accordingly, and again in full agreement to Eq.~(\ref{square_diff_const}), solenoid fringe fields change the eigen-emittances as they change the vorticity. They lower one eigen-emittance and increase the other one (see Fig.~2 of~\cite{Xiao_prstab2013} for instance). In fact the change of difference of eigen-emittances $\delta ((\Delta\epsilon )^2)$ imposed by a solenoid fringe field corresponding to a longitudinal field variation of $\delta B_s $ is according to Eqs.~(\ref{dv_eq_dW}) and (\ref{square_diff_const})
\begin{equation}
\delta ((\Delta\epsilon )^2)\,=\,A^2\,\delta\left[\frac{B_s}{(B\rho)}\right]^2\,.
\end{equation}
The same expression is obtained by doing the lengthy derivation using Eqs.~(\ref{2nd_mom_matrix},\ref{e_JMatrix},\ref{eigen12}) stated at the beginning of section~\ref{s_ebt} together with the matrix of a solenoid fringe field~\cite{Xiao_prstab2013}.

Equation~(\ref{D_vort}) can be very conveniently applied to short beam line elements along which the beam rms-area~$A$ is approximated as constant. This is the case for solenoid fringe fields as well as for tapered foils for instance. The latter were proposed for tailoring eigen-emittances by inducing longitudinal to horizontal coupling~\cite{Mccrady}. For short elements the change of difference of eigen-emittances simplifies to
\begin{equation}
\label{eq_short_elements}
\delta((\Delta\varepsilon )^2)\,=\,A^2(m_{41}-m_{23})^2\,.
\end{equation}
	
\subsection{Rigidly rotating object}
\label{rot_object}
\begin{figure}[hbt]
	\centering
	\includegraphics*[width=88mm,clip=]{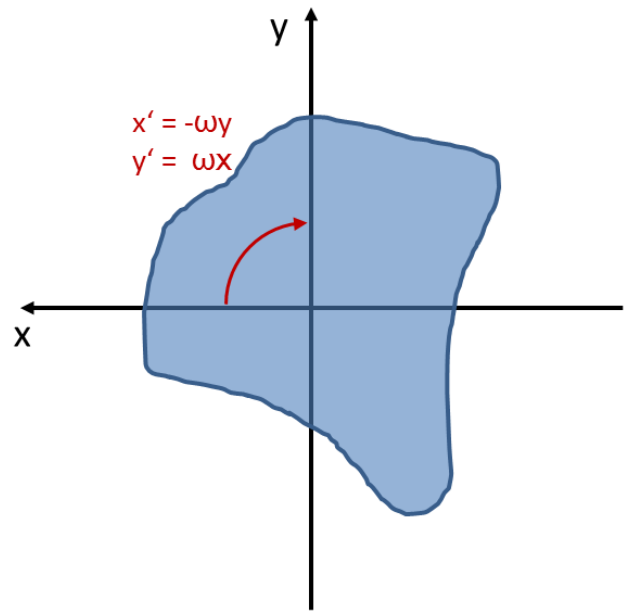}
	\caption{Arbitrary object rotating rigidly with a constant angular velocity $\omega$ around the beam axis.}
	\label{f_rotation}
\end{figure}

Equation~(\ref{square_diff_const}) is used to calculate the eigen-emittance of a rigidly rotating two dimensional object depicted in Fig.~\ref{f_rotation}.
On this object the particle coordinates are $(x,-\omega y,y,\omega x)$, where the unit of $\omega$ is $1/m$ since $s$ is varied along the beam line. Prior to rotation the object was at rest and both eigen-emittances as well as its vorticity were equal to zero. Rotation by $\omega$ causes the vorticity $2\omega$ and in order to satisfy Eq.~(\ref{square_diff_const}) it causes a difference of eigen-emittances of $2|\omega |A$ with $A$ as rms-area (see Eq.~(\ref{e_Arms})). Applying Eqs.~(\ref{2nd_mom_matrix},\ref{e_JMatrix},\ref{eigen12}) reveals $\varepsilon_1=2|\omega |A$ and \mbox{$\varepsilon_2=0$} confirming the above result.

\subsection{Shearing object}
\begin{figure}[hbt]
	\centering
	\includegraphics*[width=88mm,clip=]{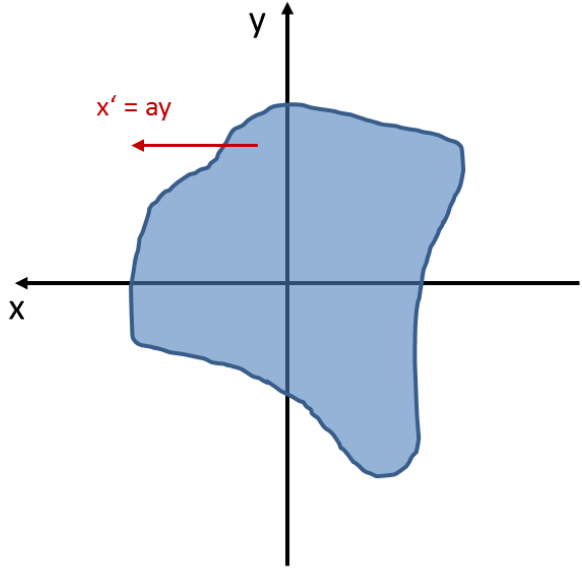}
	\caption{Arbitrary area at a constant horizontal shear with strength~$a$.}
	\label{f_shearing}
\end{figure}

In an analogue way the eigen-emittances of a shearing object depicted in Fig.~\ref{f_shearing} are calculated. The coordinates are $(x,ay,y,0)$ which by applying Eq.~(\ref{square_diff_const}) results in the eigen-emittance difference of $|a|A$. The same result is obtained by using Eqs.~(\ref{2nd_mom_matrix},\ref{e_JMatrix},\ref{eigen12}), namely $\varepsilon_1=|a|A$ and~$\varepsilon_2=0$. 

\subsection{Beams with cylindrical symmetry}
Beams with cylindrical symmetry feature $\langle xy\rangle =0$, $\langle x^2\rangle=\langle y^2\rangle=A$, and $\epsilon_{rms} := \epsilon_x = \epsilon_y $. Kim~\cite{Kim} has shown that in this case
\begin{equation}
\label{e_Kim_or}
\epsilon_{1/2}\,=\,\epsilon_{rms}\pm\mathcal{L} 
\end{equation}
with $2\mathcal{L}:=\langle xy'\rangle-\langle x'y\rangle $ and accordingly
\begin{equation}
\label{e_de_eq_2L}
\epsilon_1-\epsilon_2\,=\,\Delta\epsilon\,=\,2\mathcal{L}\,.
\end{equation}
Using Eqs.~(\ref{rms_Vorticity}) and~(\ref{e_Wa}) delivers
\begin{equation}
\langle y^2\rangle\langle xy'\rangle - \langle x^2\rangle\langle x'y\rangle\,=\,A \,\int\limits_A \,\left[\vec{\nabla}\times\vec{\bar{r'}}\right] \cdot d\vec{A}\,,
\end{equation}
which by exploiting the cylindrical symmetry properties stated above together with Eq.~(\ref{e_de_eq_2L}) gives
\begin{equation}
\Delta\epsilon\,=\,2\mathcal{L}\,=\,\langle xy'\rangle - \langle yx'\rangle\,=\,\int\limits_A \,\left[\vec{\nabla}\times\vec{\bar{r'}}\right] \cdot d\vec{A}\,,
\end{equation}
which after taking the square of both sides corresponds to Eq.~(\ref{square_diff_const}) with the constant being equal to zero. This had to be proven for cylindrical symmetric beams.

Equation~(\ref{e_Kim_or}) is a special case of Eq.~(\ref{square_diff_const}) with $const$=0 and $\mathcal{W}=2\mathcal{L}$. In fact, for round objects rotating with $\omega$=$\omega (r)$ the relation $\mathcal{W}=2\mathcal{L}$ applies, which is not the case for non-round objects. Accordingly, Eq.~(\ref{square_diff_const})  is a more general form of Eq.~(\ref{e_Kim_or}), with the first being valid also for beams without cylindrical symmetry and the role of angular momentum $2\mathcal{L}$ is taken over by the beam phase integral~$\mathcal{W}$.

\subsection{Tracking of a non-symmetric coupled beam}
Finally an initially uncoupled beam has been rms-tracked using linear transport matrices through a beam line comprising coupling elements such as solenoids and skewed quadrupoles. Initial beam parameters are listed in Tab.~\ref{tab_beamparams} and the beam line elements are listed in~Tab.~\ref{tab_beamline}.
\begin{table}[hbt]
	\caption{Initial beam parameters of the tracking calculations.}
	\begin{tabular}{l|c}
		Parameter & Value\\
		\hline
		kin. energy & 11.45~MeV/u\\
		mass number & 14\\
		charge number & 4\\
		$\varepsilon _{x}$ & 4.0 mm~mrad\\
		$\varepsilon _{y}$ & 2.0 mm~mrad\\
		$\beta _x$ & 3.0~m\\
		$\alpha _x$ & 1.5\\
		$\beta _y$ & 1.0~m\\
		$\alpha _y$ & -2.0\\
	\end{tabular}
	\label{tab_beamparams} 
\end{table}
\begin{table}[hbt]
	\caption{Beam line of the tracking calculations: element type, length, longitudinal magnetic field, magnet field gradient, and rotation angle around the positive beam axis.}
	\begin{tabular}{l|c|c|c|c}
		Element & L [m] & B[T] & B'[T/m] & rot. angle [deg]\\
		\hline
		drift & 0.3 & 0 & 0 & 0\\
		solenoid & 0.2 & 1.0 & 0 & 0\\
		drift & 0.3 & 0 & 0 & 0\\
		hor. foc. quad. & 0.1 & 0 & 25.0 & 90\\
		drift & 0.05 & 0 & 0 & 0\\
		hor. foc. quad. & 0.1 & 0 & 20.0 & 0\\
		drift & 0.05 & 0 & 0 & 0\\
		hor. foc. quad. & 0.1 & 0 & 20.0 & 45\\
		drift & 0.05 & 0 & 0 & 0\\
		hor. foc. quad. & 0.1 & 0 & 20.0 & -45\\
		drift & 0.3 & 0 & 0 & 0\\
		solenoid & 0.55 & 0.5 & 0 & 0\\
		drift & 0.3 & 0 & 0 & 0\\
	\end{tabular}
	\label{tab_beamline} 
\end{table}

\begin{figure}[hbt]
	\centering
	\includegraphics*[width=88mm,clip=]{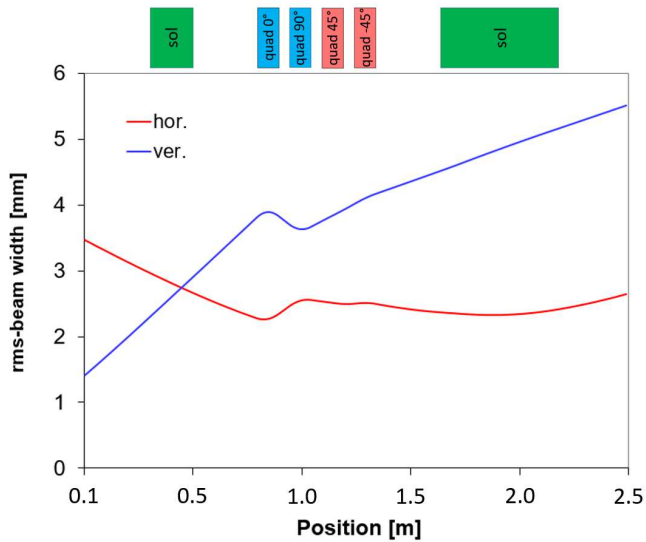}
	\caption{Horizontal (red) and vertical (blue) rms-envelope along the beam line as obtained from tracking calculations.}
	\label{f_envelopes}
\end{figure}

\begin{figure}[hbt]
	\centering
	\includegraphics*[width=88mm,clip=]{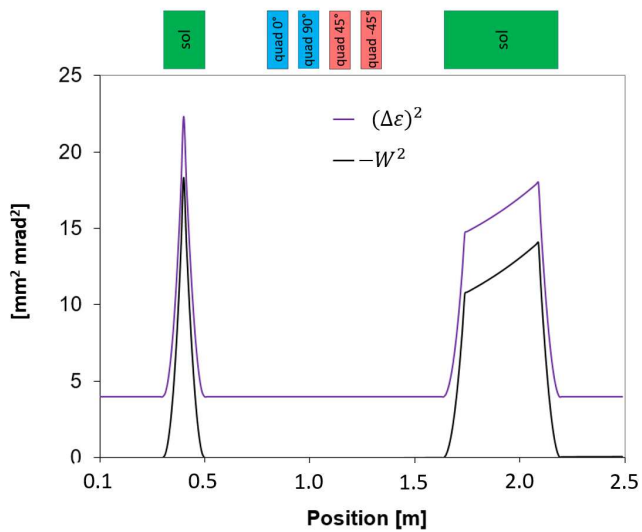}
	\caption{Squared difference of eigen-emittances (violet) and squared beam phase integral (black) as obtained from tracking calculations.}
	\label{f_dE2}
\end{figure}

\begin{figure}[hbt]
	\centering
	\includegraphics*[width=88mm,clip=]{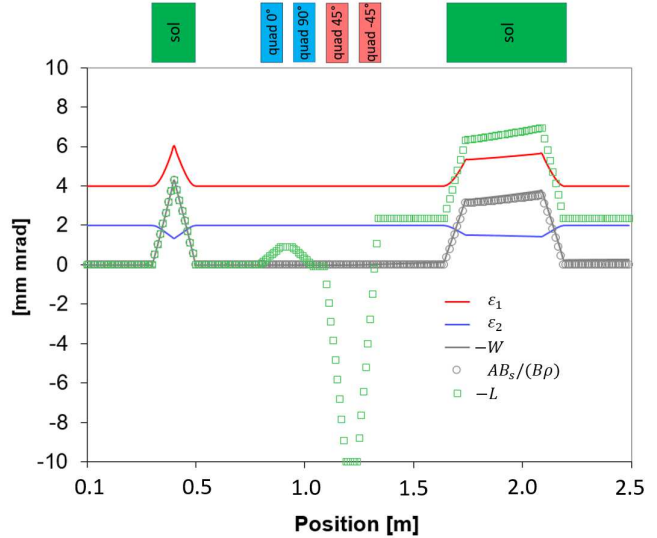}
	\caption{Eigen-emittances (red and blue lines), negative beam phase integral (gray line), $AB_s/(B\rho)$ (gray circles) and angular momentum (green boxes) as obtained from tracking calculations.}
	\label{f_W_BRho}
\end{figure}

Figures~\ref{f_envelopes} to~\ref{f_W_BRho} depict several beam parameters as functions of the position along the beam line. Rms-envelopes are shown in Fig.~\ref{f_envelopes}. They are hardly affected by the solenoids. Figure~\ref{f_dE2} compares the difference of eigen-emittances and the beam phase integral. Both vary just along regions with non-vanishing longitudinal magnetic field. The difference of their squares remains practically constant along the complete beam line. This behaviour is in agreement with~Eq.~(\ref{square_diff_const}).

Figure~\ref{f_W_BRho} plots the eigen-emittances, negative phase integral -$\mathcal{W}$, $(AB_s)/(B\rho)$, and angular momentum \mbox{$L=\langle xy'\rangle-\langle yx'\rangle =2\mathcal{L}$} along the beam line. Up to the exit of the first solenoid -$\mathcal{W}$ and $(AB_s)/(B\rho)$ are equal to each other as stated by~Eq.~(\ref{v_p_W_const}). Afterwards there is slight augmentation of~-$\mathcal{W}$ starting behind the first solenoid. This augmentation is from the finite length of the solenoid for which its exit fringe field does not completely remove vorticity and coupling imposed by the entrance fringe field and thus the beam is coupled behind the solenoid. Along the subsequent beam line this coupling in turn varies the phase integral $\mathcal{W}$ as described in section~\ref{sec_approx}. It must be stressed that this increase of vorticity occurs after the eigen-emittance tailoring itself has been accomplished inside the solenoid. Real scenarios sufficiently remove coupling and hence vorticity immediately after having tailored the eigen-emittances.
For these reasons the mentioned increase shown in Fig.~\ref{f_W_BRho} has no impact on quantitative modeling of the tailoring,  which therefore is done correctly. The tailoring is made through the fringe regions and not through the main field region. Accordingly, real scenarios keep the solenoid main region as short as possible. 

Figure~\ref{f_W_BRho} also emphasizes the difference between beam path integral $\mathcal{W}$ and angular momentum~$L$. Both are equal to each other up to the first quadrupole. Afterwards these two quantities differ significantly and the absolute value of angular momentum exceeds the one of the beam phase integral. The difference of eigen-emittances is given by the beam path integral rather than by the angular momentum. This is an important statement of Eq.~(\ref{square_diff_const}) and an essential difference to Eq.~(\ref{e_Kim_or}). The latter assigns this role to angular momentum and applies just to beams with cylindrical symmetry. The appendix gives two illustrative examples on the difference of angular momentum and beam phase integral and their role on eigen-emittances. 

Finally, it is noted that albeit $\delta A\approx\,$0 is not fulfilled along the complete line, Fig.~\ref{f_dE2} demonstrates that Eq.~(\ref{square_diff_const}) holds very well. The next two subsections are on modeling of eigen-emittance variation performed in experiments.

\subsection{Flat beam experiment at FERMILAB}
At FERMILAB's NICCAD photoinjector an electron beam has been created on a cathode surface of area~$A_0$ being immersed into a longitudinal magnetic field of~$B_0$~\cite{Piot_prstab2006}. The beam has been extracted into a region without magnetic field, accelerated, and transversely decoupled. Applying Eqs.~(\ref{v_p_W_const}) and (\ref{square_diff_const}), the final difference of normalized eigen-emittances can be calculated very quickly.

Since the beam was born with $\Delta\varepsilon _i$=0 and $\mathcal{W}_i$=0, the constant of Eq.~(\ref{square_diff_const}) is equal to zero. Hence applying this equation to the beam before and after extraction from the cathode magnetic field region gives
\begin{equation}
0\,-\,0\,\,=\,\,(\Delta\varepsilon _f)^2\,-\,\mathcal{W}_f^2\,,
\end{equation}
i.e., $\Delta\varepsilon _f=\mathcal{W}_f$. This is used in Eq.~(\ref{v_p_W_const}) being also applied to the beam before and after extraction delivering
\begin{equation}
\frac{A_0B_0}{(B\rho)}\,+\,0\,\,=\,\,0\,+\mathcal{W}_f\,\,=\,\,0\,+\,\Delta\varepsilon _f\,.
\end{equation}
To the very left and very right sides of this equation normalization of the emittances is applied and results into
\begin{equation}
\Delta\varepsilon _{nf}\,=\,\frac{A_0B_0}{(B\rho)}\beta\gamma\,=\,\frac{eA_0B_0}{m_0c}\,,
\end{equation}
where $m_0$ is the electron rest mass and $e$ is its charge. The authors of~\cite{Piot_prstab2006} used the definition \mbox{$\mathcal{L}:=(eB_0A_0)/(2m_0\beta\gamma c)$} and hence one obtains for the final difference of eigen-emittances
\begin{equation}
\label{eq_Piot}
\Delta\varepsilon _{nf}\,=\,2\beta\gamma\mathcal{L}
\end{equation}
being exactly the result stated in~\cite{Piot_prstab2006} and fully equivalent to~Eq.~(\ref{e_de_eq_2L}). It is stressed that Eq.~(\ref{eq_Piot}) has been obtained from Eqs.~(\ref{v_p_W_const}) and (\ref{square_diff_const}) without requiring a round beam, while Eq.~(\ref{e_de_eq_2L}) has been derived by explicitly requiring a round beam~\cite{Kim}.

\subsection{Emittance transfer experiment at GSI}
At GSI transverse eigen-emittances have been tailored by passing a nitrogen beam through a short solenoid~\cite{groening_prl}. A stripping foil has been placed at the solenoid center. Doing so, the ions passed the entrance and exit fringe region with different charge state and the imposed vorticities did not compensate each other to zero. Since the exit charge state~(7+) has been larger than the entrance charge state~(3+), the beam vorticity was effectively changed as through a stand-alone exit fringe field passed with the charge state of~4+. Accordingly, the two eigen-emittances have been changed as well. Decoupling of the planes has been performed shortly behind the solenoid by a skewed quadrupole triplet. The parameters of the experiment are listed in Tab.~\ref{tab_emtex}.
\begin{table}[hbt]
	\caption{Beam parameters of the emittance transfer experiment~EMTEX~\cite{groening_prl} in front of and behind the solenoid.}
	\begin{tabular}{l|c}
		Parameter & Value\\
		\hline
		kin. energy & 11.45~MeV/u\\
		mass number & 14\\
		$\varepsilon _{x,3+}$ & 1.040 mm~mrad\\
		$\varepsilon _{y,3+}$ & 0.825 mm~mrad\\
		$A_f$ & 4.166 mm$\mathrm{^2}$\\
		$q_{in/out}$ & 3+ / 7+\\
		$q_{eff}$ & 4+\\
		$(B\rho )_{3+/7+/4+}$ & 2.278 / 0.976 / 1.710 Tm\\
		$B_f$ & 0.9 T\\
	\end{tabular}
	\label{tab_emtex} 
\end{table}

Calculation of the final difference of eigen-emittances compares the parameters in front of the effective stand-alone exit fringe field, i.e., at the stripping foil right after stripping ($B_s$=$B_f$=0.9~T) to those after the solenoid~($B_s$=0). Beginning with Eq.~(\ref{v_p_W_const}) gives
\begin{equation}
\frac{A_fB_f}{(B\rho)_{eff}}\,+\,0\,\,=\,\,0\,+\,\mathcal{W}_{7+}\,,
\end{equation}
i.e., $\mathcal{W}_{7+}=(A_fB_f)/(B\rho)_{eff}$. Using this result and applying Eq.~(\ref{square_diff_const}) leads to
\begin{equation}
(\Delta\varepsilon _{3+})^2-0\,=\,(\Delta\varepsilon _{7+})^2-\mathcal{W}_{7+}^2\,=\,(\Delta\varepsilon _{7+})^2-\left[\frac{A_fB_f}{(B\rho)_{eff}}\right]^2
\end{equation}
and to the final difference of eigen-emittances
\begin{equation}
\Delta\varepsilon _{7+}\,=\,\sqrt{(\Delta\varepsilon _{3+})^2+\left[\frac{A_fB_f}{(B\rho)_{eff}}\right]^2}\,.
\end{equation}
The same result is obtained by applying Eq.~(\ref{eq_short_elements}). Plugging in the values of Tab.~\ref{tab_emtex} together with \mbox{$\Delta\varepsilon _{3+}$=(1.04-0.825)~mm~mrad} delivers
\begin{equation}
|\Delta\varepsilon _{7+}|\,=\,\mathrm{2.203~mm~mrad}\,,
\end{equation}
being in very good agreement with the measured value of~2.0(4)~mm~mrad.

\section{Eigen-emittances and rotations}
\label{s_eig_rot}
This section shows that any beam can be regarded as being equivalent to two objects with different areas performing rigid rotations with angular velocities~$\pm\omega$ around the beam axis~(under the assumptions stated in section~\ref{sec_approx}). To this end we initially refer to the original work of R.~Brinkmann~\cite{Brinkmann_rep}. Therein a beam extracted from a cathode being immersed into a solenoidal field is considered. Its phase space coordinates are those of an object rotating with $\omega$, i.e.,
\begin{equation}
\begin{bmatrix}
x_0\\
x_0'\\
y_0\\
y_0'\\
\end{bmatrix}
\,=\,
\begin{bmatrix}
x_0\\
-\omega y_0\\
y_0\\
\omega x_0\\
\end{bmatrix}\,.
\end{equation}
This beam can be transformed into a flat beam inhabiting just horizontal phase space coordinates differing from zero. The according transformation $\mathcal{P}_x(\omega)$ is from a beam line segment $M(I,\pi /2)$, being an identity in the horizontal phase space and providing $\pi /2$ of phase advance in the vertical space space. The segment  $M(I,\pi /2)$ is rotated by $\pi /4$ around the positive beam axis delivering~\cite{Brinkmann_rep}
\begin{equation}
\mathcal{P}_x(\omega)\,:=\,R(\frac{\pi}{4})
\begin{bmatrix}
1 & 0 & 0 & 0\\
0 & 1 & 0 & 0\\
0 & 0 & 0 & 1/\omega \\
0 & 0 & -\omega & 0 \\
\end{bmatrix}
\,R(-\frac{\pi}{4}) \\
\end{equation}
or
\begin{equation}
\label{e_def_projector}
\mathcal{P}_x(\omega)\,=\,\frac{1}{2}
\begin{bmatrix}
1 & -1/\omega & 1 & 1/\omega \\
\omega & 1 & -\omega & 1\\
1 & 1/\omega & 1 & -1/\omega \\
-\omega & 1 & \omega & 1\\
\end{bmatrix}\,.
\end{equation}
Accordingly,
\begin{equation}
\mathcal{P}_x(\omega)
\begin{bmatrix}
x\\
-\omega y\\
y\\
\omega x\\
\end{bmatrix}
\,=\,\frac{1}{2}
\begin{bmatrix}
2x+2y\\
2\omega x - 2\omega y\\
0\\
0\\
\end{bmatrix}
\,:=\,
\begin{bmatrix}
u\\
u'\\
0\\
0\\
\end{bmatrix}\,,
\end{equation}
being a flat beam with just horizontal dimensions. These projectors were the base for the round-to-flat transformation suggested and demonstrated by~\cite{Brinkmann_rep} and~\cite{Edwards}. They have the properties
\begin{equation}
\mathcal{P}_x^{-1}(\omega)\,=\,\mathcal{P}_x(-\omega)
\end{equation}
\begin{equation}
\mathcal{P}_x^2(\omega)\,=\,
\begin{bmatrix}
0 & 0 & -1 & 0\\
0 & 0 & 0 & -1\\
-1 & 0 & 0 & 0\\
0 & -1 & 0 & 0\\
\end{bmatrix}\,,
\end{equation}
i.e., reflection at $x+y=0$ and $x'+y'=0$ and hence
\begin{equation}
\mathcal{P}_x^4(\omega)\,=\,I\,.
\end{equation}

In the following we use the consideration that if $\mathcal{P}_x(\omega)$ projects a rotating object into an oscillation in one single transverse plane, application of $\mathcal{P}_x^{-1}(\omega)= \mathcal{P}_x(-\omega)$ should be the inverted process, that means projection of two transverse oscillations onto two rotating objects.

Any arbitrary distribution may be regarded as a combination of oscillations in the two transverse planes. Application of $\mathcal{P}_x^{-1}(\omega)$ to arbitrary particle coordinates gives
\begin{equation}
\mathcal{P}_x(-\omega)
\begin{bmatrix}
x\\
x'\\
y\\
y'\\
\end{bmatrix}
\,=\,\frac{1}{2}\left[
\begin{bmatrix}
x+x'/\omega\\
-\omega x+x'\\
x-x'/\omega\\
\omega x+x'\\
\end{bmatrix}
+
\begin{bmatrix}
y-y'/\omega\\
\omega y+y'\\
y+y'/\omega\\
-\omega y+y'\\
\end{bmatrix}
\right]\,.
\end{equation}
By defining
\begin{equation}
	\nonumber
a\,:=\,x+\frac{x'}{\omega}\,,
\end{equation}
\begin{equation}
	\nonumber
b\,:=\,x-\frac{x'}{\omega}\,,
\end{equation}
\begin{equation}
	\nonumber
c\,:=\,y-\frac{y'}{\omega}\,,
\end{equation}
\begin{equation}
	\nonumber
d\,:=\,y+\frac{y'}{\omega}\,,
\end{equation}
the previous expression can be re-stated as
\begin{equation}
\mathcal{P}_x(-\omega)
\begin{bmatrix}
x\\
x'\\
y\\
y'\\
\end{bmatrix}
\,=\,\frac{1}{2}
\begin{bmatrix}
a\\
-\omega b\\
b\\
\omega a\\
\end{bmatrix}
+\frac{1}{2}
\begin{bmatrix}
c\\
\omega d\\
d\\
-\omega c\\
\end{bmatrix}\,,
\end{equation}
describing a superposition of an object $(ab)$ rotating with $\omega$ and an object $(cd)$ rotating with $-\omega$.

Calculation of the eigen-emittances corresponding to these two objects is straight forward. As these rotating objects have $\epsilon_{4d}:=\epsilon_1\epsilon_2=0$ one eigen-emittance is equal to zero and the positive difference of eigen-emittances is equal to the other eigen-emittance. Applying the findings from subsection~\ref{rot_object}, the non-zero eigen-emittances of the two objects are
\begin{equation}
\label{e_epsab}
\epsilon_{ab}\,=\,2|\omega| A_{ab}\,\,\,\,\mathrm{and}\,\,\,\,\epsilon_{cd}\,=\,2|\omega| A_{cd}\,,
\end{equation}
with $A_{ab}$ and $A_{cd}$ being the rms-areas of these objects. It is left to determine $A_{ab}$ and $A_{cd}$ using the definitions of $a,b,c,d$ and Eq.~(\ref{e_Arms}). Doing so for $A_{ab}$ one obtains
\begin{equation}
\begin{split}
16A_{ab}^2\,& =\,\langle a^2\rangle\langle b^2\rangle - \langle ab\rangle ^2 \\
& \,=\frac{4}{\omega ^2}\langle x^2\rangle\langle x'^2\rangle-\frac{4}{\omega ^2}\langle xx'\rangle \\
& \,=\frac{4}{\omega ^2}\epsilon_x ^2\,.
\end{split}
\end{equation}
Accordingly, the areas of the objects are simply linked to the beam rms-areas through
\begin{equation}
\label{e_Aab}
A_{ab}\,=\,\frac{\epsilon_x}{2|\omega|}\,\,\,\,\mathrm{and}\,\,\,\,A_{cd}\,=\,\frac{\epsilon_y}{2|\omega|}\,.
\end{equation}
Comparison of Eqs.~(\ref{e_epsab}) and (\ref{e_Aab}) reveals that the eigen-emittances of the two objects are equal to the transverse beam rms-emittances
\begin{equation}
\label{e_eigen_rms}
\epsilon_{ab}\,=\epsilon_x,\,\,\,\mathrm{and}\,\,\,\,\epsilon_{cd}\,=\epsilon_y\,.
\end{equation}
This result has been obtained very rapidly by assuming validity of Eq.~(\ref{square_diff_const}). It can be confirmed by going through the arduous deviation of plugging the definitions of $a,b,c,d$ into Eqs.~(\ref{2nd_mom_matrix},\ref{e_JMatrix},\ref{eigen12}).

\section{Conclusions and outlook}
The quantity $\mathcal{W}$ has been introduced as the beam phase integral or the integrated beam vorticity. For cases of practical interest it was shown that change of difference in transverse eigen-emittances is given by the change of the beam phase integral. This is a generalization of previous findings that assigned this difference to angular momentum for the special case of cylindrical symmetric beams, for which the beam phase integral and angular momentum merge. However, for non-symmetric beams the two quantities are different and the eigen-emittance difference is given by the beam phase integral instead.

The new relation drastically simplifies calculations of eigen-emittances as well as quantitative modeling of experiments that tailor beam eigen-emittances.
Additionally, it allows for gaining an improved physical picture of the nature of eigen-emittances. Beams are eqivalent to superposition of two rotating objects rotating with $\pm\omega$ inhabiting rms-areas being equal to the beam transverse rms-emittances devided by $2|\omega|$.

As final remark we notice the occurrence and relevance of phase integrals in the frame of beam emittance variation through longitudinal magnetic fields. Recently the variation of single particle angular momentum in such environments has been revisited with the perspective of quantization of this angular momentum~\cite{floettmann_pra,karlovets_njp}. Future works may address whether this concept could be extended to beam emittances.

\appendix
\section{Illustration of difference between beam angular momentum and phase integral}
This paragraph shall illustrate the difference between angular momentum
\begin{equation}
\nonumber
L\,=\,\langle xy'\rangle - \langle yx'\rangle
\end{equation}
and beam phase integral
\begin{equation}
\nonumber
\mathcal{W}\,=\,\frac{1}{A}\left[\langle y^2\rangle\langle xy'\rangle - \langle x^2\rangle\langle yx'\rangle + \langle xy\rangle (\langle xx'\rangle - \langle yy'\rangle)\right]\,,
\end{equation}
with $A=\,\sqrt{\langle x^2\rangle\langle y^2\rangle -\langle xy\rangle ^2}$ together with their role in the context of eigen-emittances.
The first example treats an ellipse performing two different types of rotation and the second is on a thin wire moving along a rectangular orbit.

\subsection{Rotating ellipse}
\begin{figure}[hbt]
	\centering
	\includegraphics*[width=88mm,clip=]{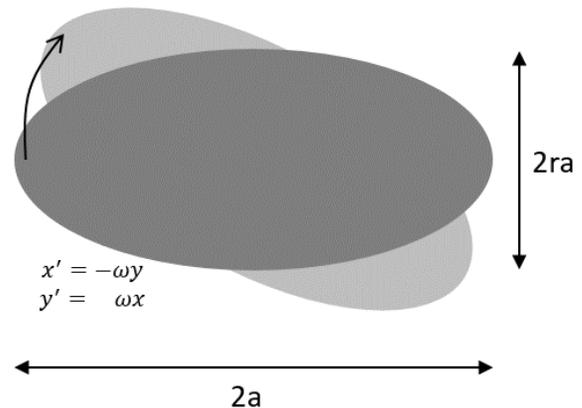}
	\caption{Ellipse performing rigid and concentric rotation.}
	\label{f_rigrot}
\end{figure}

The ellipse depicted in Fig.~\ref{f_rigrot} starts a rigid rotation around its center and accordingly its second moments $\langle x^2\rangle$, $\langle y^2\rangle$, and $\langle xy\rangle$ vary during rotation. $L_{rig}$, $\mathcal{W}_{rig}$, and the two eigen-emittances $\epsilon _{1/2,rig}$ remain constant. Calculation of these four quantities from the second moments delivers
\begin{equation}
\nonumber
L_{rig}\,=\,\frac{\omega}{4}a^2(1+r^2)\,,
\end{equation}
\begin{equation}
\nonumber
\mathcal{W}_{rig}\,=\,\frac{\omega}{2}a^2r\,,
\end{equation}
\begin{equation}
\nonumber
\epsilon_{1,rig}\,=\,\Delta\epsilon\,=\,\mathcal{W}_{rig}\,\leq\,L_{rig}\,,
\end{equation}
\begin{equation}
\nonumber
\epsilon_{2,rig}\,=\,0\,.
\end{equation}
Here the eigen-emittances were calculated by applying Eqs.~(\ref{2nd_mom_matrix},\ref{e_JMatrix},\ref{eigen12}).

\begin{figure}[hbt]
	\centering
	\includegraphics*[width=88mm,clip=]{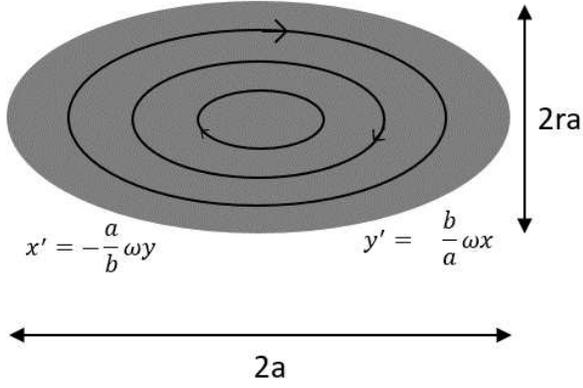}
	\caption{Ellipse performing intrinsic and concentric rotation.}
	\label{f_introt}
\end{figure}

Instead in the case depicted in Fig.~\ref{f_introt} the ellipse performs an intrinsic rotation preserving its second moments $\langle x^2\rangle$, $\langle y^2\rangle$, and $\langle xy\rangle$. Re-calculation of $L$, $\mathcal{W}$, and $\epsilon_{1/2}$ results into
\begin{equation}
\nonumber
L_{int}\,=\,\frac{\omega}{2}a^2r\,,
\end{equation}
\begin{equation}
\nonumber
\mathcal{W}_{int}\,=\,\frac{\omega}{4}a^2(1+r^2)\,,
\end{equation}
\begin{equation}
\nonumber
\epsilon_{1,int}\,=\,\Delta\epsilon\,=\,\mathcal{W}_{int}\,\geq\,L_{int} \,,
\end{equation}
\begin{equation}
	\nonumber
	\epsilon_{2,int}\,=\,0\,.
\end{equation}
Comparison of rigid and intrinsic rotation reveals that swapping the type of rotation swaps the expressions for $L$ and $\mathcal{W}$. Nonetheless, for both types the difference of eigen-emittances~$\Delta\varepsilon$ is given by the beam path integral $\mathcal{W}$ and not by the angular momentum~$L$. Just for the special case of cylindrical symmetry $r$=1 the two types of rotation merge as do the two quantities, i.e., \mbox{$L=\mathcal{W}=\omega a^2/2$}.

\subsection{Moving wire}
\begin{figure}[hbt]
	\centering
	\includegraphics*[width=88mm,clip=]{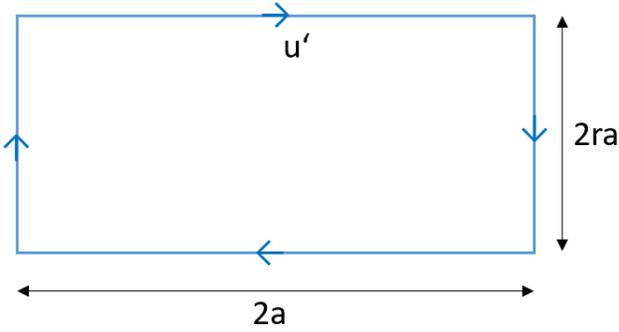}
	\caption{Thin wire moving around a rectangular area.}
	\label{f_wirefig}
\end{figure}

\begin{figure}[hbt]
	\centering
	\includegraphics*[width=88mm,clip=]{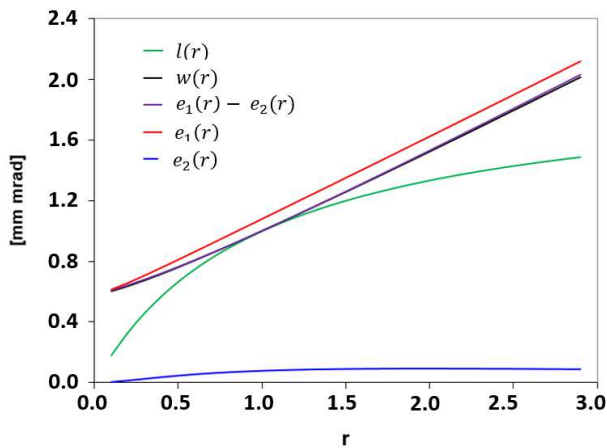}
	\caption{Angular momentum (green), beam phase integral or vorticity (black), difference of eigen-emittances (violet), and the two eigen-emittances (red and blue) as functions of the geometric aspect ratio~$r$.}
	\label{f_wireplot}
\end{figure}

The second example considers a thin wire starting to move with constant velocity $u'$ around a rectangular area as depicted in~Fig.~\ref{f_wirefig}.
Calculation of its acquired angular momentum, beam phase integral or vorticity, and eigen-emittances is straight forward. It is lengthy for the eigen-emittances using Eqs.~(\ref{2nd_mom_matrix},\ref{e_JMatrix},\ref{eigen12}) and it delivers
\begin{equation}
	\nonumber
	L\,=\,au'\cdot l(r)
\end{equation}
\begin{equation}
	\nonumber
	\mathcal{W}\,=\,au'\cdot w(r)
\end{equation}
\begin{equation}
	\nonumber
	\varepsilon_{1/2}\,=\,au'\cdot e_{1/2}(r)\,,
\end{equation}
with
\begin{widetext}
	\begin{equation}
		\nonumber
		l(r)\,=\,\frac{2r}{1+r}
	\end{equation}
	\begin{equation}
		\nonumber
		w(r)\,=\,\frac{(1+r)^2}{\sqrt{3r^2+10r+3}}
	\end{equation}
	\begin{equation}
		\nonumber
		e_{1/2}\,=\,\frac{1}{\sqrt{6}(1+r)}\sqrt{(1+r)^4-r(r^2+1)\pm\sqrt{(1+r)^8-2r(r+1)^4(r^2+1)+r^2(r^2-1)^2}}\,.
	\end{equation}
\end{widetext}
As for the ellipse these quantities depend on the geometric aspect ratio~$r$ and the $r$-dependence of the angular momentum $l(r)$ is different w.r.t. the one of the beam phase integral and the difference of eigen-emittances. While the two latter scale proportional to $r$ for large~$r$, the angular momentum converges to~$l(r)$=2. At first glance the expressions of $w(r)$ and of the difference of eigen-emittances seem different, but they are actually almost identical. Figure~\ref{f_wireplot} plots the angular momentum, the phase integral, and the eigen-emittances as functions of the aspect ratio~$r$.
It reveals that the difference of eigen-emittances is very well approximated by the phase integral (the relative difference is less than 1\%). Comparing $w(r)$ to $e_{1/2}(r)$, the derivation of $w(r)$ is significantly faster and the final expression is much simpler than the one for~ $e_{1}(r)-e_2(r)$.

The angular momentum instead is obviously systematically off from the difference of eigen-emittances. Just for the special case of symmetry~$r$=1, the phase integral or vorticity is equal to the angular momentum, i.e., $l(1)$=$w(1)$=$e_1(1)-e_2(1)$=1. Hence, another evidence has been provided that change of difference of eigen-emittances is generally given by change of vorticity and not by change of angular momentum.


\begin{thebibliography}{99}
\bibitem{rot_inv}
It can be shown that $\mathcal{W}$ from Eq.~(\ref{rms_Vorticity}) is invariant under rotation around the beam axis.
\bibitem{safe_ass}
It is re-stressed that these statements refer to the laboratory coordinates $(x,y,x',y')$ being relevant for applications. In conjugate coordinates the eigen-emittances are preserved also along solenoid main and fringe fields. This preservation is the base of the extended Busch theorem~\cite{groening_prab}.
\end{thebibliography}
\end{document}